\documentclass{ws-procs9x6}

\begin{document}

\title{ATLAS SILICON MICROSTRIP TRACKER: \\ OPERATION AND PERFORMANCE}

\author{V.~A.~MITSOU, for the ATLAS SCT Collaboration}

\address{Instituto de F\'isica Corpuscular (IFIC), CSIC -- Universitat
  de Val\`encia, P.O.~Box~22085, E-46071, Valencia, Spain \\
E-mail: vasiliki.mitsou@ific.uv.es}

\begin{abstract}
The Semiconductor Tracker (SCT) is a silicon strip detector and one of the key precision tracking devices in the Inner Detector of the ATLAS experiment at CERN LHC. The completed SCT has been installed inside the ATLAS experimental cavern since 2007 and has been operational since then. Calibration data has been taken regularly and analyzed to determine the performance of the system. In this paper the current status of the SCT is reviewed, including results from data-taking periods in 2010 and 2011. We report on the operation of the detector including overviews on services, connectivity and observed problems. The main emphasis is given to the performance of the SCT with the LHC in collision mode and to the performance of individual electronic components. 
\end{abstract}


\bodymatter

\section{Introduction}

The ATLAS detector\cite{Aad:2008zzm}, one of the two general-purpose experiments at the Large Hadron Collider (LHC) at CERN, has been taking proton-proton collision data at a centre-of-mass energy of 7~TeV since March 2010.  The ATLAS Inner Detector (ID)\cite{Aad:2010bx} combines silicon detector technology (pixels\cite{Aad:2008zz} and microstrips\cite{Ahmad:2007zza}) in the innermost part with a straw drift detector\cite{Abat:2008zza} with transition radiation detection capabilities (Transition Radiation Tracker,
TRT) on the outside, operating in a 2-T superconducting solenoid.

\begin{figure}
\begin{center}
\psfig{file=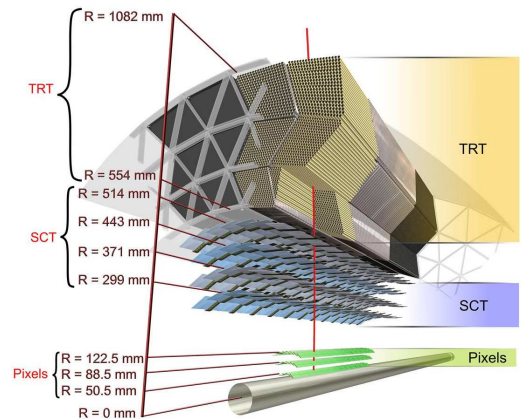,width=0.6\linewidth}
\end{center}
\caption{Layout of the ATLAS Inner Detector: it comprises the Transition
    Radiation Detector, the Semiconductor Tracker and the Pixel system.}
\label{fig:ID}
\end{figure}

The microstrip detector (Semiconductor Tracker, SCT), as shown in Fig.~\ref{fig:ID}, forms the middle layer of the ID between the Pixel detector and the TRT. The SCT system comprises a barrel\cite{Abdesselam:2008zzd} made of four nested cylinders and two end-caps\cite{Abdesselam:2008zzc} of nine disks each. The barrel layers carry 2112~detector units (modules)\cite{Abdesselam:2006wt} altogether, while a total of 1976~end-cap modules\cite{Abdesselam:2007ec} are mounted on the disks. The whole SCT occupies a cylinder of 5.6~m in length and 56~cm in radius with the innermost layer at a radius of 27~cm. 

The silicon modules\cite{Abdesselam:2006wt,Abdesselam:2007ec} consist of one or two pairs of single-sided p-\emph{in}-n microstrip sensors glued back-to-back at a 40-mrad stereo angle to provide two-dimensional track reconstruction. The $285$-$\mu{\rm m}$ thick sensors\cite{Ahmad:2007zza} have 768 AC-coupled strips with an $80~\mu{\rm m}$ pitch for the barrel and a \mbox{$57-94~\mu{\rm m}$} pitch for the end-cap modules. Barrel modules follow one common design, while for the forward ones four different types exist according to their position in the detector. The readout\cite{Abdesselam:2008zza} of each module is based on 12~ABCD3TA ASICs\cite{Campabadal:2005rj} manufactured in the radiation-hard DMILL process mounted on a copper/kapton hybrid. Each module is designed, constructed and tested to operate as a stand-alone unit, mechanically, electrically, optically and thermally. 

\section{Operation stability}

\begin{figure}
\begin{center}
\psfig{file=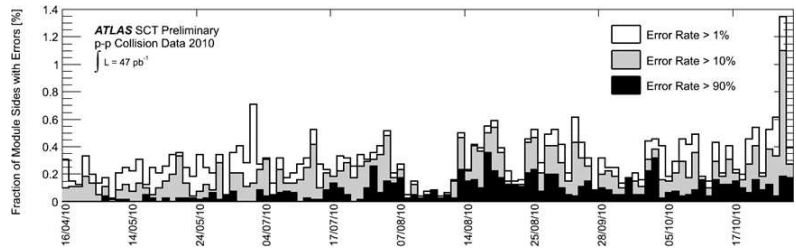,width=0.91\linewidth}
\end{center}
\caption{Fraction of SCT module sides that are reporting errors as a function of time (run number) for the data taken in 2010.}
\label{fig:error1}
\end{figure}

The performance of the SCT modules and substructures have been repeatedly tested in the past in dedicated beam tests (before 2004), during the Combined Test Beam with other ID components in summer 2004 and with cosmic rays\cite{Abat:2008zzd} both on the surface and after installation in the ATLAS cavern. Since autumn 2009 various aspects of the SCT operation have been studied continuously while recording physics data in $pp$ collisions at 900~GeV, 2.36~TeV and, since March 2010, at 7~TeV. The SCT performs very well with increasing collision rate, as demonstrated in Fig.~\ref{fig:error1}, where the fraction of SCT module sides giving errors as a function of time is shown. It is stressed that the time interval spans an increasing instantaneous luminosity of five orders of magnitude. The SCT data-quality inefficiency is mainly due to HV ramping up during LHC stable-beam declaration, with an overall 99.9\% operation efficiency. The total fraction of data with SCT errors remained less than 0.25\% throughout 2010. 

The only issues encountered affecting the --- otherwise stable --- SCT configuration are: (a) a faulty connection to a cooling loop\cite{Attree:2008zz} discovered during commissioning, that cannot be repaired since it is located behind the endplate; and (b) some unexpected failures\cite{Cooke:2011nm} of off-detector optical transmitters (TX-plugins)\cite{Abdesselam:2007zz} that are being replaced by humidity-resistant plugins. In total, more than 99\% of all SCT modules are fully operational.

\section{SCT performance} 

\begin{figure}
\begin{center}
\psfig{file=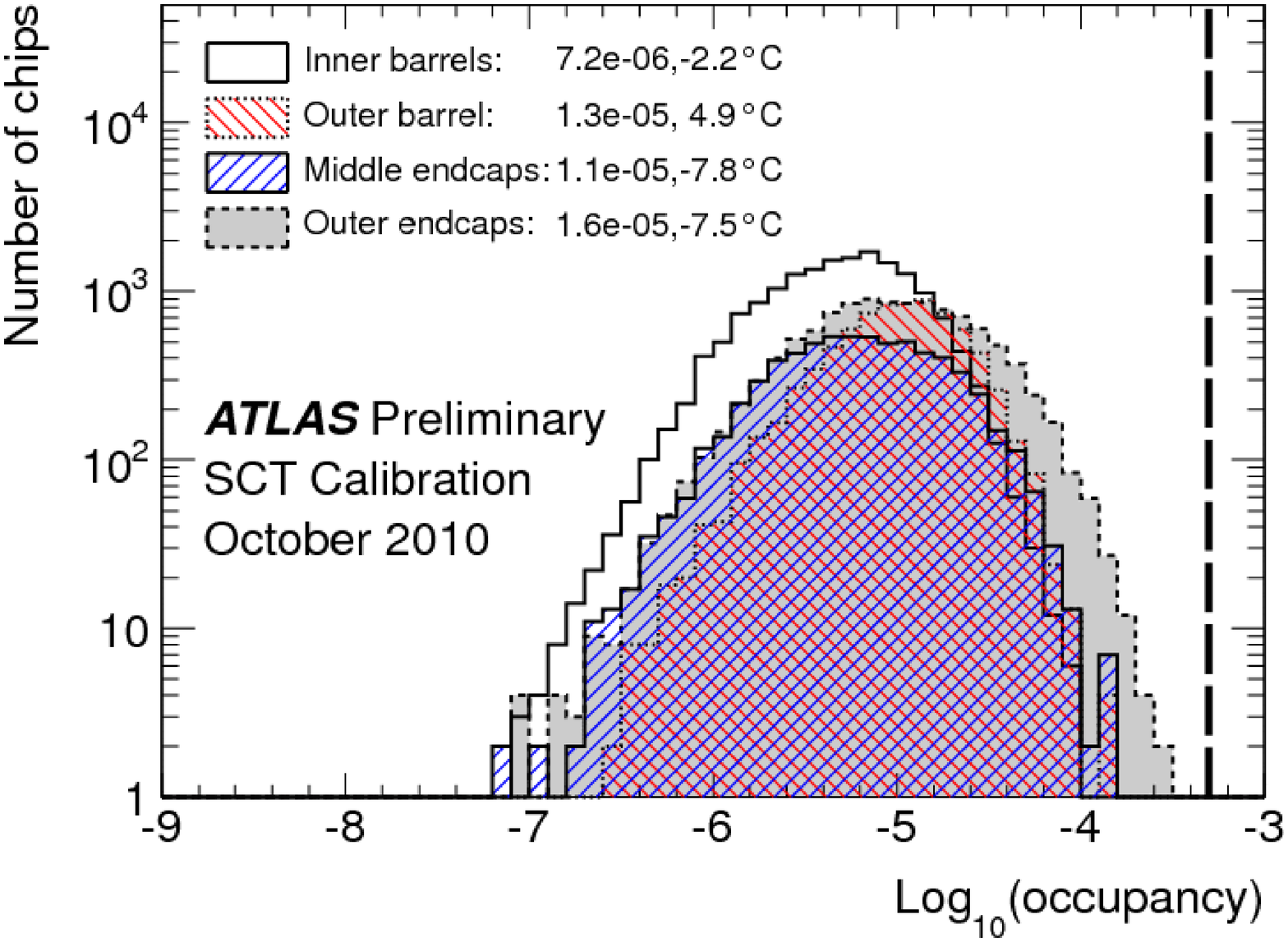,width=0.45\linewidth}\hfill
\psfig{file=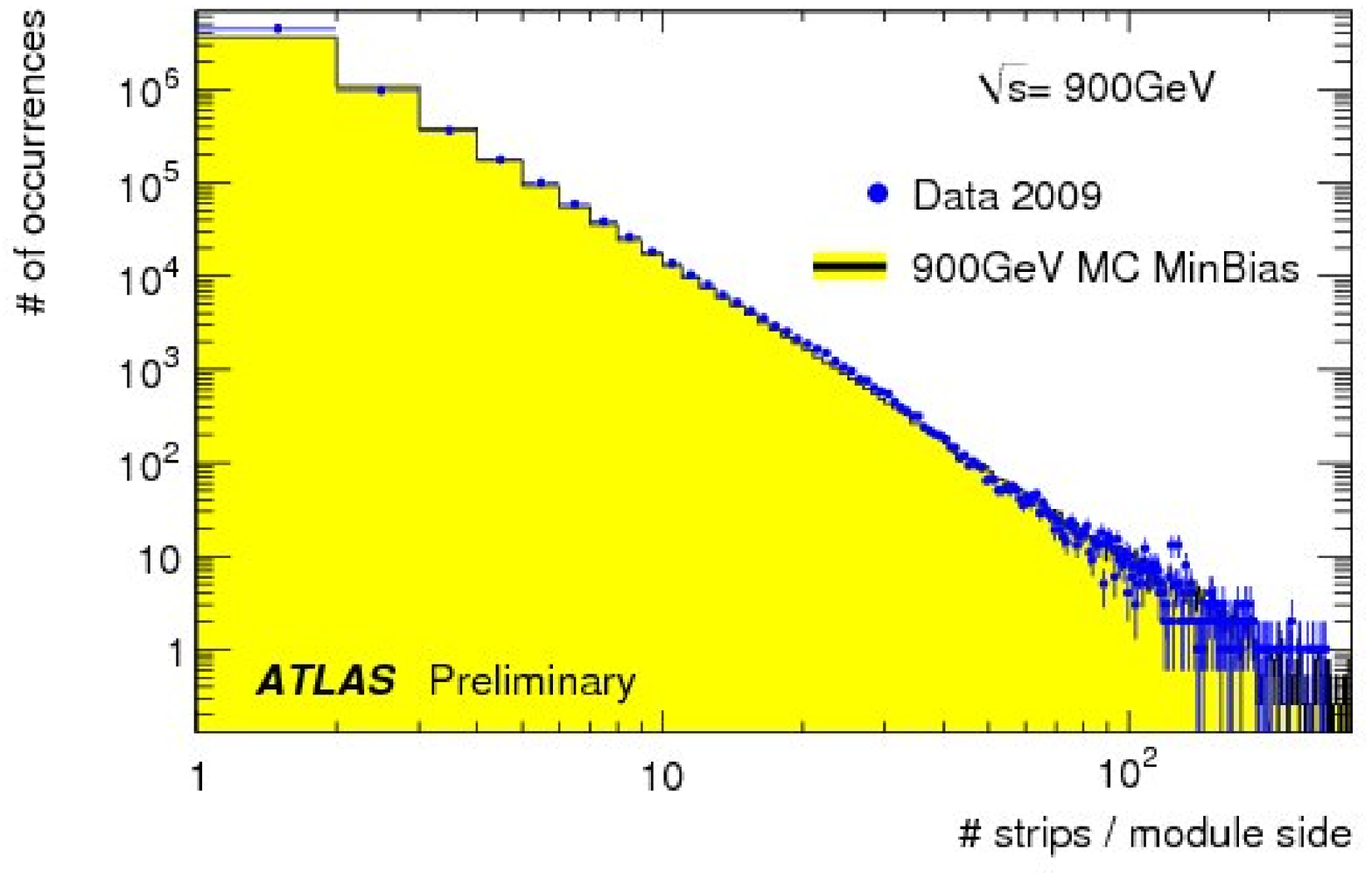,width=0.5\linewidth}
\end{center}
\caption{Left: Distributions of noise occupancy for each chip split up according to the module type. Right: Number of hit strips per module side for 900~GeV data compared with a minimum bias Monte Carlo sample.}
\label{fig:NO-strips}
\end{figure}

Noise measurements are performed by two methods:\cite{Mitsou:2006xc} injection of calibration pulses and noise estimation from response-curve fit; and measuring noise occupancy as a function of threshold to extract the input noise. The measured values are well correlated and, as shown in Fig.~\ref{fig:NO-strips} (left), the noise occupancy is well below the required level of $5\times10^{-4}$. 

Good agreement has been observed between Monte Carlo and 900~GeV data in the number of hits per module side, as shown in Fig.~\ref{fig:NO-strips} (right). The discrepancy at low $N$ is due to the lower noise assumed in simulation compared to data by approximately a factor of three. 

As far as the SCT timing\cite{Abdesselam:2008zza} is concerned, per-module adjustments are made to account for the optical fibre length and the time-of-flight from the interaction point though timing scans. The related measurements show that the SCT is well synchronized with the Level-1 trigger.

\begin{figure}
\begin{center}
\psfig{file=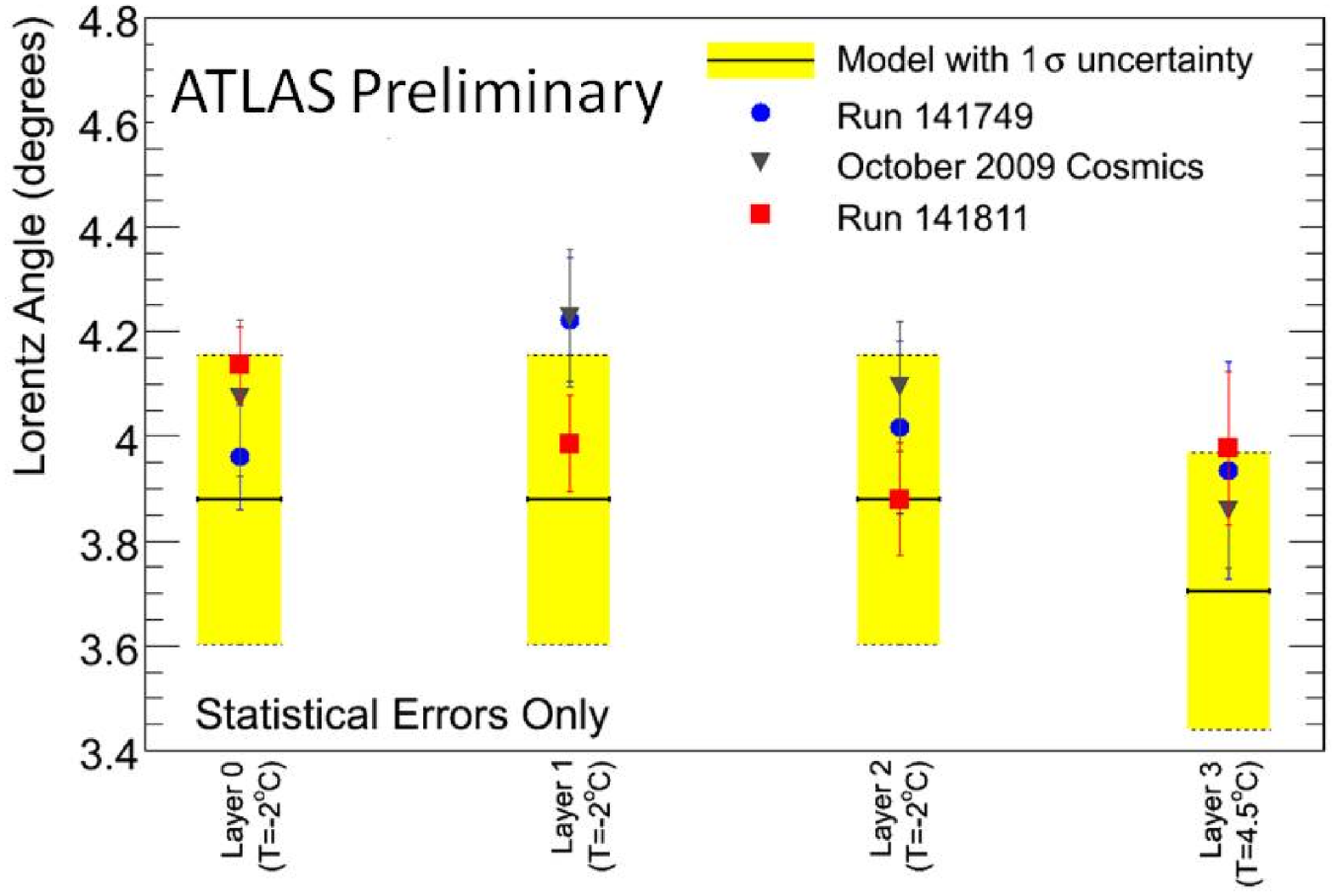,width=0.47\linewidth}\hfill
\psfig{file=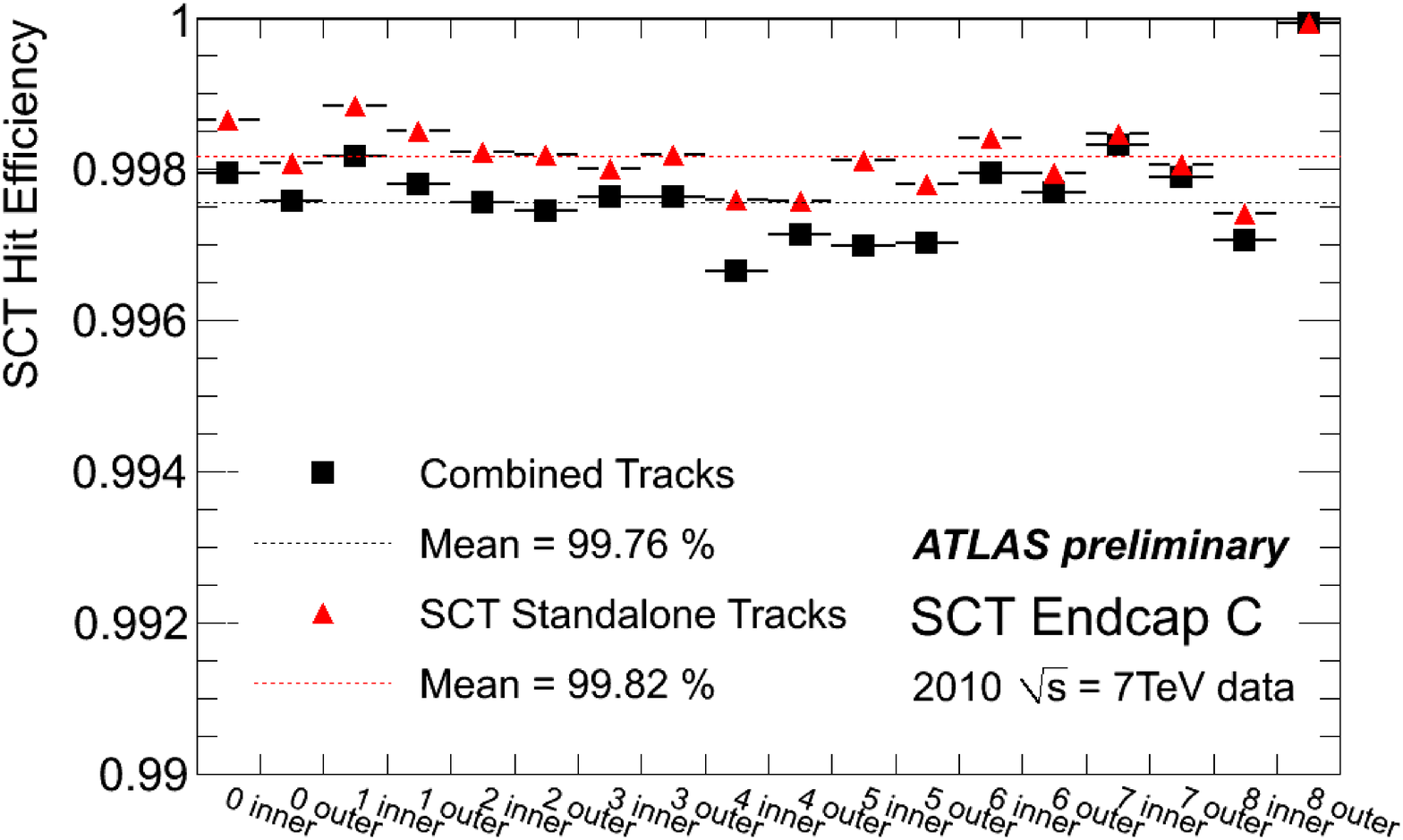,width=0.5\linewidth}
\end{center}
\caption{Left: Lorentz angle extracted from the cluster-size vs.\ angle plot for collision and cosmic ray data compared to model predictions. Right: Intrinsic module efficiency for tracks measured in the SCT end-cap C.}
\label{fig:Lorenz-eff}
\end{figure}

The Lorentz angle as determined by finding the angle corresponding to the minimum cluster width is shown in Fig.~\ref{fig:Lorenz-eff} (left) for the barrel layers and for various data/conditions. The measurements agree well with the predictions, proving the good signal digitization in the full simulation.

The intrinsic strip efficiency is computed by counting missed hits in well-reconstructed tracks with $p_\mathrm{T}>1~\mathrm{GeV}$, excluding dead modules and chips. The SCT hit efficiency for all three SCT parts (shown in Fig.~\ref{fig:Lorenz-eff}, right, for the end-cap C) is well above the 99\% specification.

\section{Radiation damage}

The radiation damage is being monitored through the leakage current. The comparison between measurements and FLUKA\cite{Battistoni:2007zzb} predictions of the fluence shows good agreement in the barrel region, whilst a discrepancy in the low-$R$ forward region is under study. The leakage current evolution with time lies within $1\sigma$ of the predicted values. The long-term monitoring of the effect during operation continues.

\section{Conclusions}

The ATLAS Semiconductor Tracker performance shows excellent performance. A total of 99.1\% of the SCT modules are used for data taking. Noise occupancy and hit efficiency are well within the design specifications. The optical transmitter failures are understood and do not impair data taking efficiency. The Monte Carlo simulation reproduces accurately the detector geometry and material budget. The radiation damage is being monitored and is in good agreement with expectations. The SCT is a key precision tracking device in ATLAS and we are taking more and more good physics data every day. The SCT commissioning and running experience is used to extract valuable lessons for future silicon strip detector projects.

\section*{Acknowledgments}
The author acknowledges support by the Spanish Ministry of Science and Innovation (MICINN) under the project FPA2009-13234-C04-01, by the Ram\'on y Cajal contract RYC-2007-00631 of MICINN and CSIC, and by the Spanish Agency of International Cooperation for Development under the PCI project A/030322/10. 


\end{document}